\documentclass[aps,pre,onecolumn,nofootinbib]{revtex4}

\usepackage{graphicx}
\usepackage{epsf}
\usepackage{epstopdf}
\usepackage{subfig}

\usepackage{amsmath}
\usepackage{amssymb}
\usepackage{mathrsfs}
\usepackage{dcolumn}

\usepackage{multirow}

\usepackage{xcolor}
\usepackage{enumerate}
\usepackage{float}
\usepackage{physics}
\usepackage{pdfpages}

\usepackage{amsthm}

\usepackage{hyperref}

\makeatletter
\newcommand*{\rom}[1]{\expandafter\@slowromancap\romannumeral #1@}
\makeatother

\def\be{\begin{equation}}
	\def\ee{\end{equation}}

\graphicspath{{images/}}

\def\be{\begin{equation}}
	\def\ee{\end{equation}}
\def\ba{\begin{eqnarray}}
	\def\ea{\end{eqnarray}}

\graphicspath{{images/}}
\begin{document}
	
	\title{Reduced-order autoregressive dynamics of a complex financial system: a PCA-based approach}
	\author{Pouriya Khalilian}
	\email{pouriya@email.kntu.ac.ir}
	\affiliation{Department of Physics, K. N. Toosi University of Technology, P.O. Box 15875-4416, Tehran, Iran}
	\author{Sara Azizi}
	\email{sarahazizi@email.kntu.ac.ir}
	\affiliation{Department of Physics, K. N. Toosi University of Technology, P.O. Box 15875-4416, Tehran, Iran}
	\author{Mohammad Hossein Amiri}
	\email{am@email.kntu.ac.ir}
	\affiliation{Department of Physics, Tehran University, Tehran, Iran}
	\author{Javad T. Firouzjaee}
	\email{firouzjaee@kntu.ac.ir}
	\affiliation{Department of Physics, K. N. Toosi University of Technology, P.O. Box 15875-4416, Tehran, Iran}
	\begin{abstract}
		This study analyzes the dynamic interactions among the NASDAQ index, crude oil, gold, and the US dollar using a reduced-order modeling approach. Time-delay embedding and principal component analysis are employed to encode high-dimensional financial dynamics, followed by linear regression in the reduced space. Correlation and lagged regression analyses reveal heterogeneous cross-asset dependencies. Model performance, evaluated using the coefficient of determination ($R^2$), demonstrates that a limited number of principal components is sufficient to capture the dominant dynamics of each asset, with varying complexity across markets.
		
	\end{abstract}

	\maketitle
	\newpage
	\tableofcontents
	
	\section{Introduction}
	
The interdependence among major economic and financial variables has long attracted
the attention of researchers, market participants, and policymakers. Over time, these
variables have undergone substantial structural and behavioral changes, making the
identification of persistent patterns and dominant interaction mechanisms increasingly
challenging. In this context, considerable effort has been devoted to understanding
the joint behavior of crude oil prices, gold prices, the US dollar, and stock market
indices, as these assets play a central role in shaping global financial conditions and
are often driven by common economic, political, and geopolitical forces
\cite{salisu2017revisiting}. From a broader perspective, financial markets may be viewed
as complex systems composed of many interacting components, where macroscopic dynamics
emerge from collective behavior and feedback effects \cite{mantegna1999introduction,
	bouchaud2003theory}.

Crude oil occupies a pivotal position in the global economy as a primary source of
energy and a key input for industrial production. Consequently, fluctuations in oil
prices have far-reaching implications for macroeconomic stability and long-term growth,
particularly in economies that are heavily dependent on oil exports or imports. While
natural resources such as oil and minerals can support economic development, the
resource curse paradox suggests that resource-rich countries often experience slower
growth compared to resource-poor economies \cite{su2021does}. Empirical evidence further
highlights a negative association between natural resource abundance and economic
performance, underscoring the complex and sometimes adverse role of commodity dependence
\cite{jianqiang2020exacerbating,naseer2020double}.

Accounting for nearly 40\% of global energy consumption, oil represents the largest
segment of the commodity market. As a result, oil price uncertainty acts as a major
source of macroeconomic instability. Historical episodes such as oil embargoes,
military conflicts, financial crises, and global pandemics have repeatedly triggered
sharp increases in oil price volatility, with significant consequences for economic
growth across countries. Gold, in contrast, is widely regarded as a financial safe
haven and a hedge against uncertainty. Periods of heightened market stress often lead
investors to rebalance their portfolios toward gold in response to inflation risks,
currency depreciation, and equity market downturns. The COVID-19 pandemic provides a
clear illustration of this behavior: while oil prices collapsed to historically low
levels, gold prices surged to record highs in 2020. Such contrasting movements suggest
that oil and gold price dynamics are closely linked to broader macroeconomic structures
and business cycles \cite{guan2021volatility}.

Empirical studies confirm that volatility in commodity markets has important
macroeconomic consequences. Guan et al.\ \cite{guan2021volatility} show that increased
volatility in oil and gold markets can hinder real GDP growth regardless of a country’s
exporter or importer status. At the same time, short-term increases in gold prices may
stimulate growth in gold-producing economies, whereas short-term oil price fluctuations
appear to exert a more limited impact, partly due to the long-term nature of oil
contracts. Moreover, spillover effects across markets further complicate this picture,
as fluctuations in gold prices can propagate to other financial markets with varying
intensity and timing \cite{lucey2014gold}.

The interactions among oil prices, exchange rates, gold, and stock markets have been
examined using a wide range of econometric frameworks. Arfaoui \cite{arfaoui2017oil},
for instance, employed a system of simultaneous equations to identify both direct and
indirect linkages among these assets, reporting negative relationships between oil
prices and stock markets, oil prices and the US dollar, and the US dollar and gold
prices, while also emphasizing a close connection between oil and gold. Earlier
theoretical contributions by Golub and Krugman \cite{golub1983oil,krugman1983oil}
provide economic explanations for these relationships, highlighting the role of
dollar-denominated oil pricing, portfolio allocation decisions of oil-exporting
countries, and current account dynamics.

Beyond linear economic relationships, increasing evidence suggests that commodity and
financial markets exhibit nonlinear, feedback-driven, and potentially chaotic
dynamics. The growing financialization of oil markets has strengthened their
connections with other asset classes, while the inherent nonlinearity of oil price
time series poses substantial challenges for conventional modeling approaches. To
address these issues, a variety of advanced methods have been proposed, including
machine learning and chaos-based models \cite{karasu2022crude}, as well as nonlinear
time-series analyses providing empirical evidence of chaotic structures in oil price
dynamics \cite{ling2011chaotic}. From a dynamical systems perspective, such behavior can
be naturally described using time-delay embedding techniques, which allow the
reconstruction of the underlying system dynamics from observed time series
\cite{takens1981detecting,kantz2004nonlinear}.

Gold markets exhibit similarly complex behavior. Although gold is often perceived as a
stable store of value, its price dynamics are influenced by a combination of inflation
expectations, exchange rate movements, monetary policy, and broader economic and
political conditions. The dynamic relationships among gold prices, exchange rates, oil
prices, and stock market returns have been investigated using vector autoregression and
cointegration techniques, revealing stable yet time-dependent interactions among these
variables \cite{sujit2011study}. These observations further reinforce the view that
financial markets form a tightly coupled system characterized by multiscale
interactions and feedback mechanisms \cite{alados2014positive}.

From the perspective of statistical mechanics, financial markets may be viewed as
high-dimensional, non-equilibrium systems composed of interacting agents. Price
dynamics emerge from collective behavior, feedback mechanisms, and stochastic
fluctuations, giving rise to complex temporal correlations and emergent structures.
Within this framework, dimensionality reduction techniques provide a natural approach
for identifying the dominant modes that govern system dynamics and for separating
informative signals from noise \cite{laloux1999noise,brunton2016discovering}. Motivated
by this view, the present study adopts a reduced-order modeling strategy based on
dynamically encoded time series and principal component analysis to investigate the
joint dynamics of oil, gold, the US dollar, and the NASDAQ index, aiming to extract the
essential structure of market interactions while filtering out redundant and noisy
components.

	\section{Qualitative Analysis of Financial Market Dynamics}
	This section provides a qualitative analysis of the main financial variables considered in this study, including crude oil, gold, and the NASDAQ index. Rather than focusing on formal modeling or quantitative estimation, the aim here is to outline the economic and dynamical roles of these variables and to summarize the key mechanisms through which they interact with one another. By reviewing empirical findings and established observations from the literature, this section highlights the structural relationships, feedback effects, and nonlinear features that characterize financial market behavior. This qualitative perspective serves as a conceptual foundation for the quantitative framework developed in the subsequent sections.
	
	\subsection{Crude Oil}
	Crude oil is a cornerstone of the global economy, serving as a primary energy source and a critical input for industrial production. Consequently, fluctuations in oil prices exert widespread effects across economic sectors. Empirical evidence indicates that oil price volatility imposes costs on virtually all sectors, reflecting the strong dependence of many economies on oil production, trade, and revenues \cite{an2019oil}. This dependence has motivated extensive efforts to forecast oil prices, as oil price expectations directly influence investment decisions and macroeconomic planning \cite{nyangarika2019oil}.
	
	Oil price movements affect inflation, stock market valuations, and economic growth in both oil-importing and oil-exporting countries \cite{an2019oil}. In particular, changes in oil prices propagate through supply chains, transportation costs, and consumer prices, thereby influencing daily economic activity \cite{nyangarika2019oil,firouzjaee2022lstm}. These effects extend beyond direct energy consumption, impacting sectors such as aviation, automotive industries, and long-term procurement strategies that rely on oil price forecasts \cite{nyangarika2019oil}.
	
	A key channel linking oil to global financial markets is the US dollar, as most oil contracts are denominated in dollars \cite{sujit2011study}. Dollar depreciation reduces oil-export revenues and raises production costs for exporting countries, while higher oil prices tend to appreciate exporters’ currencies and weaken those of importers \cite{sujit2011study}. Empirical studies further show a strong inverse relationship between oil prices and the dollar exchange rate, with a 1\% dollar depreciation associated with a more than proportional increase in oil prices \cite{novotny2012link}. This mechanism dampens oil price shocks in non-dollar economies and highlights the systemic role of exchange rates.
	
	Beyond monetary factors, oil prices are shaped by supply-side constraints and nonlinear production dynamics. Delays in expanding extraction and refining capacity introduce nonlinearities that increase price sensitivity as capacity limits are approached \cite{dees2008assessing}. While refinery utilization affects oil prices, evidence suggests that expanding refinery capacity alone does not necessarily lower prices, underscoring the complexity of oil market dynamics \cite{dees2008assessing}. Geopolitical events also contribute to oil price uncertainty. Although earlier studies suggest limited effects of military conflicts on oil prices \cite{nyangarika2019oil}, recent crises such as the Russia--Ukraine war reveal that geopolitical shocks can trigger substantial short-term volatility \cite{firouzjaee2022machine}.

	\subsection{Gold}
	Gold occupies a distinctive position in financial markets as a globally recognized store of value and a hedge against macroeconomic uncertainty. Unlike country-specific assets, gold is largely independent of national economic structures and is widely perceived as protection against inflation and financial instability \cite{shen2014us}. During periods of economic stress, empirical evidence consistently shows rising gold prices alongside declining stock market valuations, reinforcing gold’s role as a safe-haven asset \cite{arfaoui2017oil}.
	
	Gold prices are closely linked to movements in the US dollar. Increases in the dollar index typically depress gold prices, while dollar depreciation supports gold appreciation \cite{shen2014us}. This inverse relationship is particularly pronounced during episodes of inflationary pressure and stock market downturns, when investors reallocate portfolios toward gold \cite{arfaoui2017oil}. Studies further indicate a positive relationship between gold prices and energy prices, while financial market indicators exhibit a negative association with gold returns \cite{juarez2011applying}.
	
	The interaction between gold prices and macroeconomic variables extends to inflation dynamics. Gold price shocks have persistent inflationary effects in developing economies and shorter-lived impacts in developed economies \cite{oloko2021fractional}. Moreover, gold prices contain forward-looking information about future inflation, especially in countries operating under explicit inflation-targeting regimes \cite{tkacz2007gold}. These findings highlight gold’s dual role as both a financial hedge and a macroeconomic signal.

	\subsection{NASDAQ}
	The NASDAQ is one of the world’s largest stock exchanges and plays a central role in global equity markets. Market microstructure and price formation mechanisms on NASDAQ have attracted considerable academic attention, particularly regarding inefficiencies surrounding market opening and closing prices \cite{pagano2008quality}. The introduction of call auctions was motivated by concerns over unreliable price discovery and aimed to improve fairness and efficiency in the trading process.
	
	Opening and closing prices serve critical functions in valuation, market marking, and empirical research, making their stability essential \cite{pagano2013call}. Continuous trading links orders sequentially, often generating short-term volatility, whereas call auctions aggregate orders and execute trades at a single price, reducing transient inefficiencies \cite{pagano2013call}. Empirical evidence shows that the introduction of call auctions reduced bid--ask spreads and volatility, particularly around market open and close, with heterogeneous effects across firm sizes \cite{pagano2013call}.
	
	Despite their benefits, auction-based mechanisms are influenced by trader coordination and participation incentives. Studies report higher failure rates for opening and closing calls in small-cap stocks, suggesting that auctions are not universally optimal \cite{ellul2005opening}. Theoretical and empirical analyses attribute these outcomes to coordination motives, whereby traders condition their participation on expectations about others’ behavior \cite{pagano1989trading,chakraborty2012order}. Overall, NASDAQ price dynamics reflect a complex interaction of microstructural rules, trader behavior, and broader market conditions, motivating reduced-order and multivariate approaches to capture their dominant drivers.

	\subsection{Financial Time Series as Chaotic Systems}
	Forecasting financial time series remains a challenging task due to their intrinsic non-stationarity, high volatility, and nonlinear structure. As noted in Ref.~\cite{kazem2013support}, stock market prediction is particularly difficult because financial series do not exhibit constant statistical properties over time. Similarly, Ref.~\cite{qi2020event} emphasizes that financial transaction data are complex, highly volatile, and inherently non-stationary.
	
	The pronounced volatility observed in many economic sectors motivates the development of more reliable modeling and forecasting approaches \cite{juarez2011applying}. In this context, complexity and chaos-based models provide a powerful framework for analyzing nonlinear phenomena, uncovering hidden dependencies among variables, and mathematically characterizing evolving market behavior \cite{juarez2011applying}. Empirical evidence suggests that several financial indices exhibit signatures consistent with chaotic dynamics, reinforcing the inadequacy of purely linear modeling approaches.
	
	Crude oil prices offer a prominent example of such complexity. Given their central role in economic development, global stability, and geopolitics, oil prices are influenced by a wide range of interacting factors. Karasu and Altan demonstrate that the nonlinear and chaotic characteristics of crude oil time series significantly limit the predictive accuracy of conventional models \cite{karasu2022crude}. These findings support the use of advanced nonlinear and reduced-order techniques for oil price estimation and forecasting.
	
	Similar complexity arises in the joint dynamics of gold prices, exchange rates, and oil prices. Ref.~\cite{sujit2011study} concludes that no simple linear relationship exists among these variables through a common commodity channel, highlighting the intrinsically coupled and nonlinear nature of their interactions. Stock market prices further compound this complexity, as they are simultaneously influenced by oil prices, exchange rates, interest rates, foreign stock indices, and broader economic conditions \cite{park2013stock}. Although these factors may act independently, their combined effect on stock prices emerges through a dense web of interdependencies.
	
	Overall, financial markets can be viewed as high-dimensional, nonlinear, and potentially chaotic systems. This perspective motivates the adoption of dimensionality reduction and multivariate modeling strategies to extract dominant structures from noisy financial data while preserving the essential dynamics governing cross-market interactions.

	\section{PCA-Based Auto-Regression Framework}
	\label{sec:pca}
	The proposed framework is not intended as a purely predictive model, but rather
	as a reduced-order stochastic representation of the collective dynamics underlying
	a complex financial system.
	
	Within the quantitative methods adopted in this study, principal component analysis (PCA) is employed as a dimensionality reduction mechanism prior to time-series modeling. PCA is a well-established unsupervised technique that projects high-dimensional, potentially correlated variables onto a low-dimensional orthogonal subspace while preserving the dominant variance structure of the data \cite{zare2018extension,cunningham2015linear}. This transformation is particularly effective in mitigating redundancy and instability arising from correlated or irrelevant features, which commonly occur in data-rich financial time series.
	
	\subsubsection{Principal Component Projection}
	
	Let $X\in\mathbb{R}^{N\times d}$ denote the centered data matrix constructed from the lag-embedded input features. PCA seeks a low-rank representation by identifying an orthonormal basis $W_k\in\mathbb{R}^{d\times k}$ that maximizes the variance of the projected data. Formally, the projection matrix is obtained by solving
	\begin{equation}
		W_k = \arg\max_{W^\top W = I_k} \ \mathrm{Tr}(W^\top S W),
	\end{equation}
	where $S=\frac{1}{N}X^\top X$ is the empirical covariance matrix. The resulting low-dimensional representation is given by
	\begin{equation}
		Z_k = X W_k,
	\end{equation}
	which corresponds to a rank-$k$ approximation of the original feature space \cite{cunningham2015linear,jiang2011linear}. Such subspace learning enables efficient representation of high-dimensional lagged observations while retaining the most informative temporal components.
	
	\subsubsection{Autoregressive Modeling in the PCA Subspace}
	Following dimensionality reduction, the temporal dynamics of the target variable are modeled using an autoregressive process in the PCA-transformed space. Specifically, the response variable at time $t$ is expressed as
	
	\begin{equation}
		y_t = \beta_0 + \sum_{j=1}^{k} \beta_j z_{t,j} + \varepsilon_t,
	\end{equation}
	where $z_{t,j}$ denotes the $j$-th principal component at time $t$, $\{\beta_j\}_{j=1}^{k}$ are the autoregressive coefficients, and $\varepsilon_t$ is a zero-mean stochastic disturbance. This formulation can be interpreted as a reduced-order autoregressive model, where the dependence on past observations is captured through the principal components derived from lagged variables. The model parameters are estimated using ordinary least squares on the training subset, obtained via a fixed hold-out partition of the time series. By operating in the PCA subspace, the autoregressive model benefits from reduced dimensionality, alleviated multicollinearity, and improved numerical stability, which are critical for high-order autoregressive representations of financial data.
	
	\subsubsection{Quantitative Performance Assessment}
	
	The predictive performance of the proposed models is quantitatively evaluated by
	comparing the estimated values $\hat{y}_t$ with the observed realizations $y_t$ on
	an independent test set. Model accuracy is assessed exclusively through the
	coefficient of determination, $R^2$, which measures the proportion of variance in
	the target variable explained by the model.
	
	\begin{equation}
		R^2 =
		1 - \frac{\sum_{t}(y_t - \hat{y}_t)^2}
		{\sum_{t}(y_t - \bar{y})^2},
	\end{equation}
	
	where $\bar{y}$ denotes the sample mean of the observed data. The $R^2$ metric
	provides a scale-independent and interpretable measure of goodness-of-fit, making
	it particularly suitable for comparing model performance across different assets
	and dimensional configurations. In this study, $R^2$ is used consistently to
	evaluate both training and test sets, enabling the identification of optimal model
	complexity and the detection of potential overfitting effects.

	\subsubsection{Analysis over Reduced Dimensionality}
	The autoregressive modeling procedure is repeated for different numbers of retained principal components $k$. This enables a systematic investigation of the trade-off between dimensionality reduction and forecasting performance. The resulting performance trends as functions of $k$ provide quantitative evidence for selecting an appropriate reduced representation that balances temporal complexity, model parsimony, and generalization capability in financial time-series forecasting.
	\section{Results and Discussion}
	\label{Results}
	
	This section presents and discusses the empirical findings of the study by integrating qualitative economic reasoning with quantitative statistical analysis. First, the structural relationships among the Nasdaq index, crude oil, gold, and the U.S. dollar are examined through correlation analysis and lagged linear regression in order to identify the direction and relative strength of cross-asset interactions. These results are then contrasted with the qualitative insights outlined earlier, providing an empirical basis for assessing their consistency. In the second part, the predictive performance of the proposed modeling framework is evaluated using dynamically encoded features and principal component analysis, with emphasis placed on generalization behavior and dimensionality selection. Together, these results offer a comprehensive view of both the interdependence and predictability of the financial assets under consideration.
	
	\subsection{Correlation and Regression-Based Analysis of Cross-Asset Relations}
	
	\subsubsection{Correlation structure among financial variables}
	
\begin{figure}[H]
		\centering
		\includegraphics[width=0.75\linewidth]{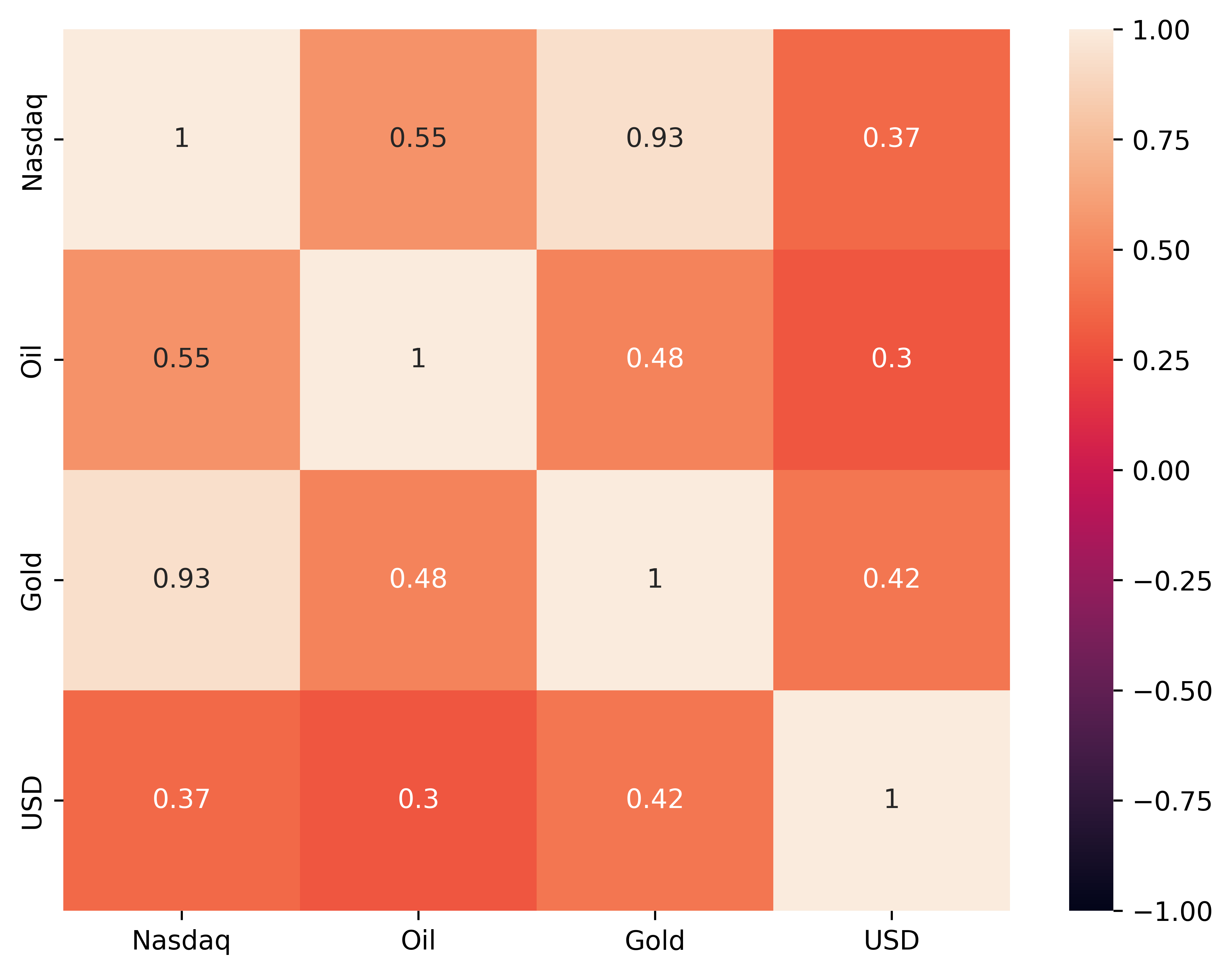}
		\caption{Pairwise Pearson correlation matrix among NASDAQ, crude oil (Oil), gold (Gold),
			and the US dollar index (USD). Color intensity represents the strength of linear
			correlation, while numerical values indicate the corresponding coefficients.}
		\label{fig:corr_heatmap}
	\end{figure}

	We begin by examining the pairwise Pearson correlation coefficients among the four
	financial variables under study: the NASDAQ index, crude oil price, gold price, and
	the US dollar index. The resulting correlation matrix is shown in
	Fig.~\ref{fig:corr_heatmap}.
	
	The correlation map reveals heterogeneous coupling strengths across the system.
	NASDAQ and gold exhibit the strongest positive co-movement ($\rho \approx 0.93$),
	while NASDAQ and oil display a moderate positive correlation ($\rho \approx 0.55$).
	Gold is moderately correlated with both oil and USD (Gold--Oil $\rho \approx 0.48$ and
	Gold--USD $\rho \approx 0.42$). In contrast, oil and USD present the weakest pairwise
	association ($\rho \approx 0.30$). These results suggest that the four-dimensional
	market state is not uniformly coupled, but rather organized into stronger and weaker
	interaction channels.
	
	While correlation analysis is informative for describing synchronous linear co-movements,
	it remains a pairwise measure and cannot determine the conditional contribution of each
	asset in the presence of the others. In particular, correlation does not distinguish
	direct from indirect associations, which motivates a multivariate regression analysis.
	
	\subsubsection{Regression-based cross-asset influence (lagged interactions)}
	
	To quantify conditional cross-asset effects, we fit a multivariate linear regression
	model for each target asset at time $t$ using the lagged values of the remaining assets
	at time $t-1$. Self-lag terms are excluded in order to isolate cross-asset interactions.
	In compact form, for each target $X \in \{\mathrm{Nasdaq},\mathrm{Oil},\mathrm{Gold},\mathrm{USD}\}$,
	\begin{equation}
		X(t)=\beta_{0}^{(X)}+\sum_{Y \neq X}\beta_{Y\rightarrow X}\,Y(t-1)+\varepsilon_{X}(t),
	\end{equation}
	where $\beta_{Y\rightarrow X}$ represents the conditional influence of asset $Y$ on $X$
	through a one-step lag.
	
\begin{figure}
	\centering
	
	\begin{minipage}{0.48\linewidth}
		\centering
		\includegraphics[width=\linewidth]{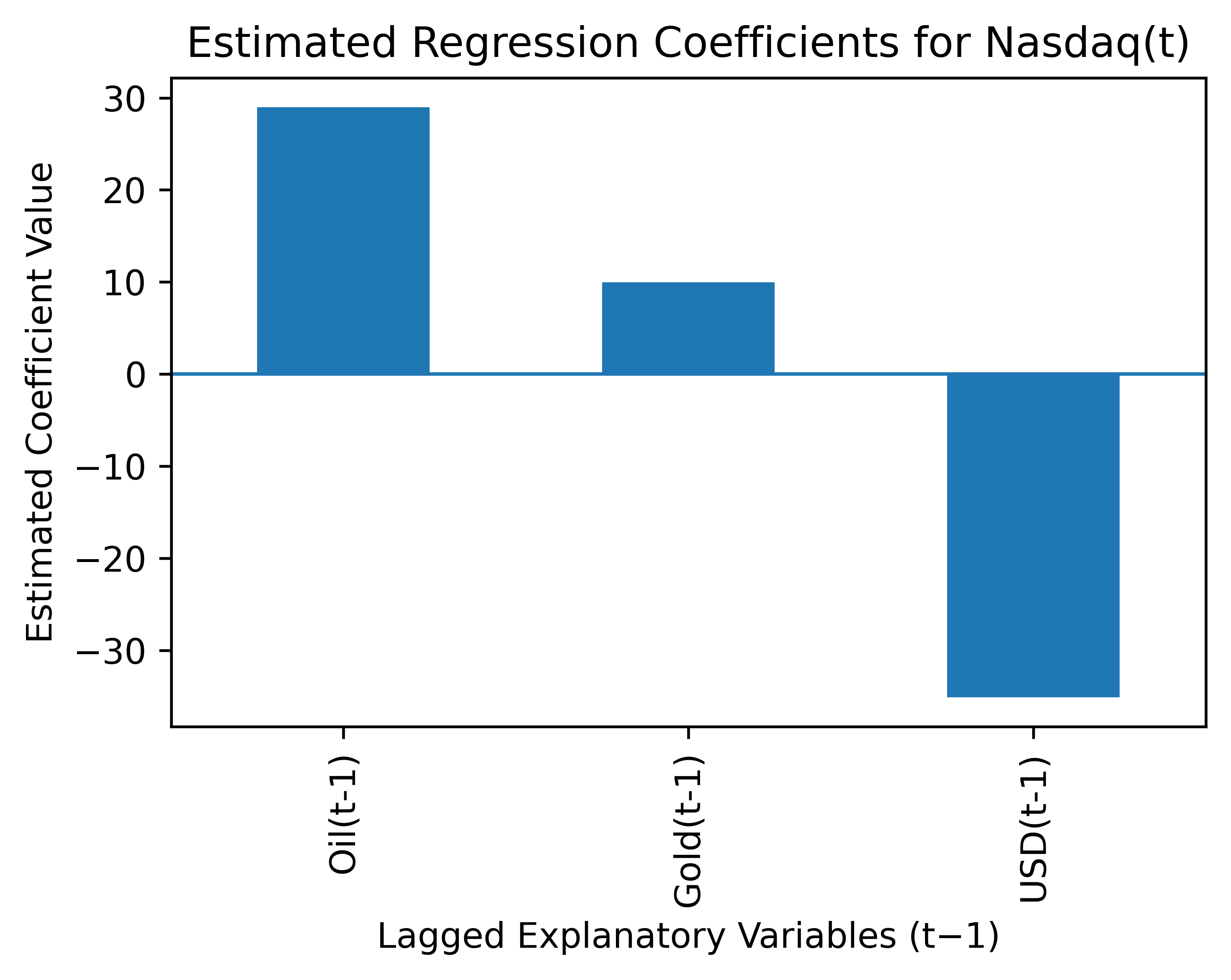}\\
		\small (a) NASDAQ as target variable.
	\end{minipage}
	\hfill
	\begin{minipage}{0.48\linewidth}
		\centering
		\includegraphics[width=\linewidth]{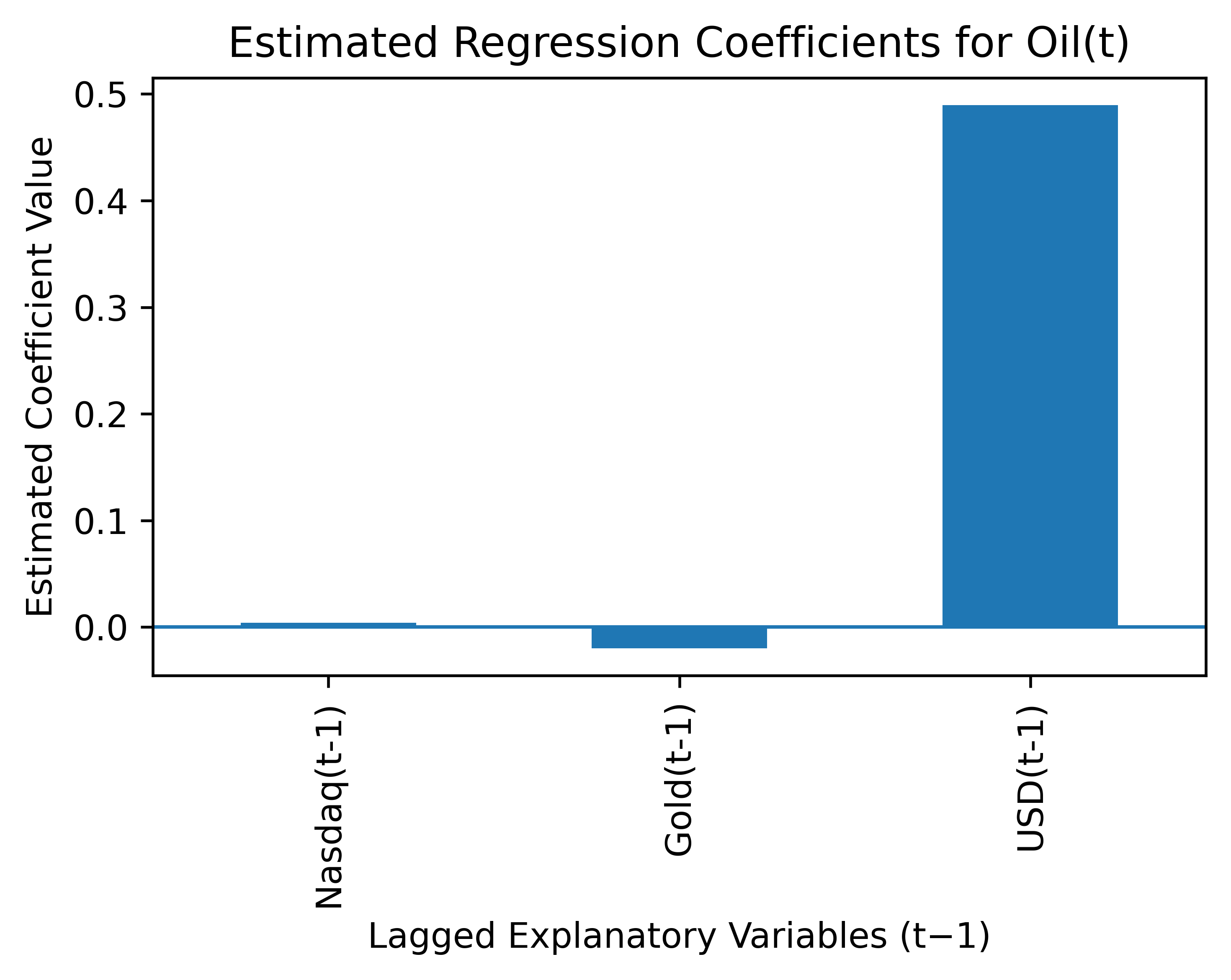}\\
		\small (b) Crude oil as target variable.
	\end{minipage}
	
	\vspace{0.8em}
	
	\begin{minipage}{0.48\linewidth}
		\centering
		\includegraphics[width=\linewidth]{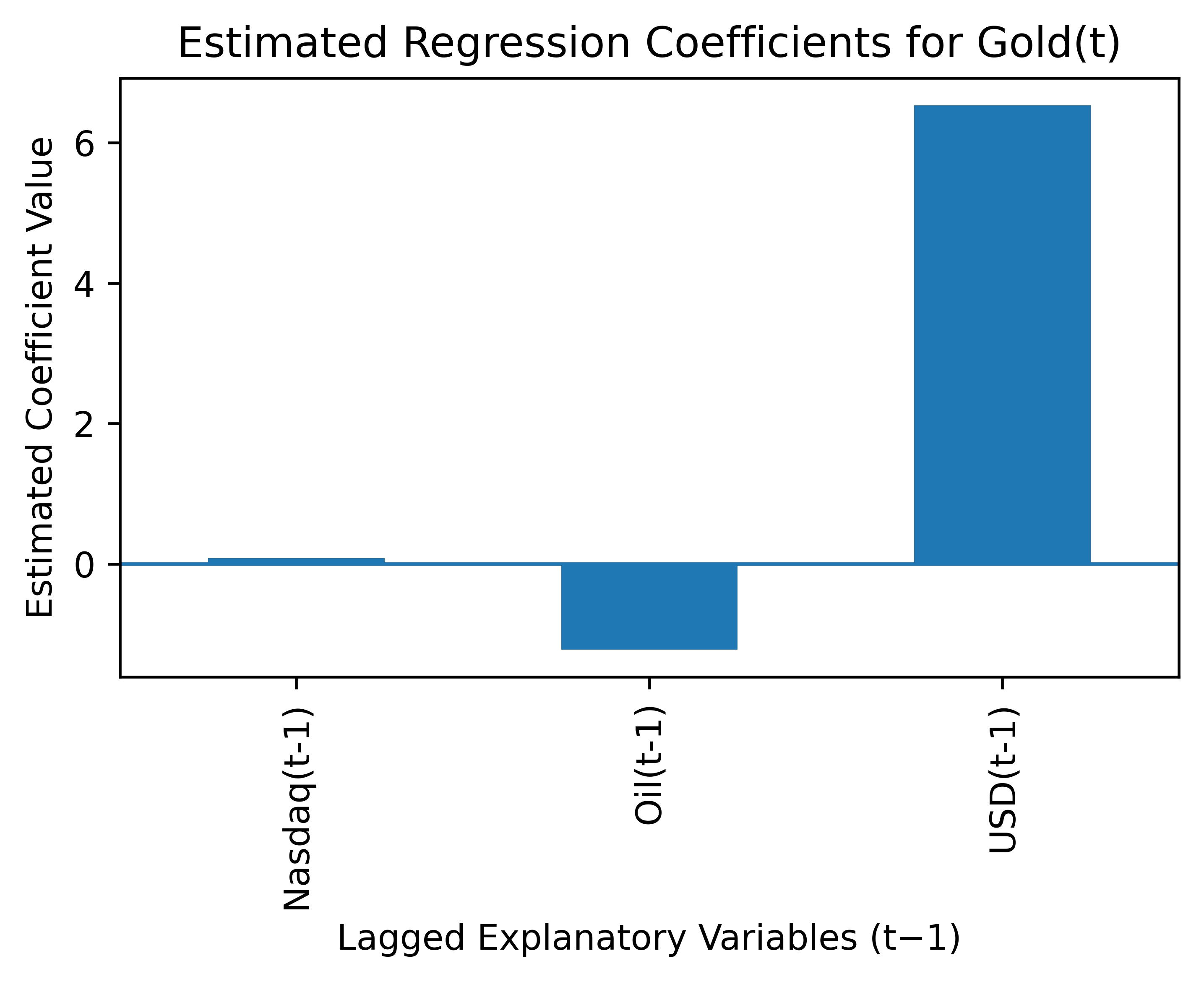}\\
		\small (c) Gold as target variable.
	\end{minipage}
	\hfill
	\begin{minipage}{0.48\linewidth}
		\centering
		\includegraphics[width=\linewidth]{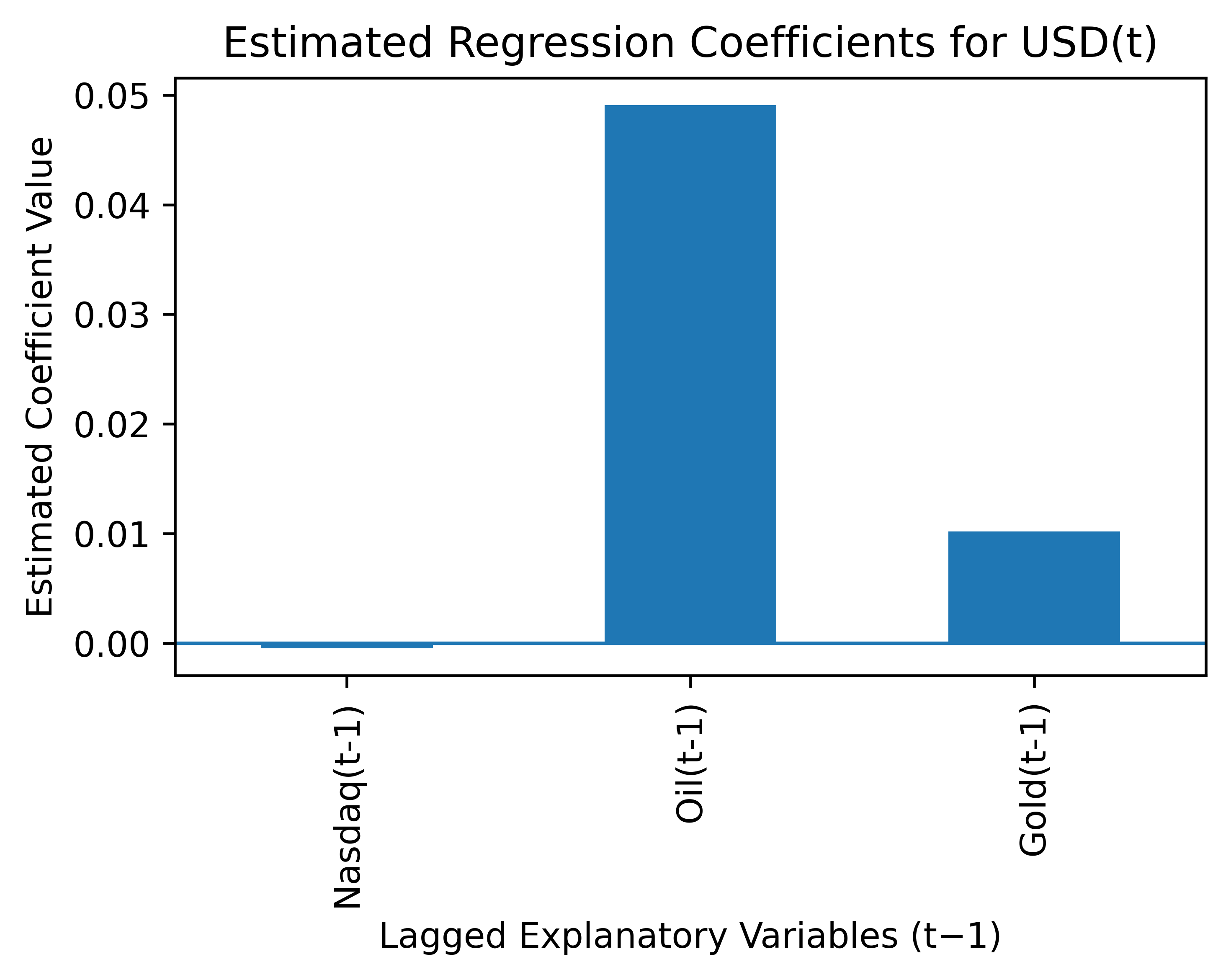}\\
		\small (d) US dollar index as target variable.
	\end{minipage}
	
	\caption{Estimated regression coefficients for the lagged multivariate linear models
		with self-lag terms excluded. Each panel shows the conditional influence of lagged
		explanatory variables at time $t-1$ on the corresponding target asset at time $t$.
		The sign and magnitude of each coefficient indicate the direction and strength of
		the cross-asset effect.}
	\label{fig:multi_reg_coef}
\end{figure}

	The estimated coefficients are visualized in Fig.~\ref{fig:multi_reg_coef}, where each
	panel corresponds to a different target variable. The numerical values are summarized
	in Table~\ref{tab:reg_coeff}.
	
	\begin{table}[H]
		\centering
		\caption{Estimated regression coefficients for the lagged multivariate linear model with self-lag terms excluded. Each row corresponds to the target variable at time $t$, and each column represents a lagged explanatory variable at time $t-1$.}
		\label{tab:reg_coeff}
		\begin{tabular}{lcccc}
			\hline
			Target variable & Nasdaq$(t\!-\!1)$ & Oil$(t\!-\!1)$ & Gold$(t\!-\!1)$ & USD$(t\!-\!1)$ \\
			\hline
			Nasdaq$(t)$ & -- & 28.96 & 10.00 & $-35.05$ \\
			Oil$(t)$    & 0.0041 & -- & $-0.0199$ & 0.4896 \\
			Gold$(t)$   & 0.0827 & $-1.2208$ & -- & 6.5347 \\
			USD$(t)$    & $-0.0005$ & 0.0491 & 0.0102 & -- \\
			\hline
		\end{tabular}
	\end{table}
	
	Taken together, Fig.~\ref{fig:multi_reg_coef} and Table~\ref{tab:reg_coeff} reveal that cross-asset interactions are strongly target-dependent and asymmetric. For example, the NASDAQ equation shows substantial positive contributions from lagged oil and gold and a large negative coefficient associated with lagged USD, suggesting that the conditional role of the dollar is dominant once other assets are controlled for. In the gold equation, USD exerts a strong positive effect, whereas oil enters with a negative coefficient, illustrating that oil--gold relations can change sign when conditioned on USD and NASDAQ. For the oil equation, the dominant coefficient is associated with USD, while the remaining effects are comparatively small. Finally, the USD equation is characterized by relatively small coefficients overall, indicating weaker one-step predictability from the other assets within this linear lagged setting.
	
	A comparison between the correlation structure in Fig.~\ref{fig:corr_heatmap} and the regression results in Fig.~\ref{fig:multi_reg_coef} highlights an important point: strong pairwise co-movement does not necessarily translate into dominant conditional influence. Correlation summarizes synchronous bivariate association, whereas regression coefficients quantify multivariate conditional effects and can reveal sign changes due to shared drivers and indirect pathways. These observations motivate the reduced-order analysis in Part~II, where collective market dynamics are examined beyond pairwise and linear summaries.
	
	\subsection{PCA-Based Dynamic Modeling and Predictive Performance}
	
	In this part, we investigate the predictive performance of linear regression models
	constructed on dynamically encoded financial time series after dimensionality
	reduction via Principal Component Analysis (PCA).
	Each asset is represented through a time-delay embedding, where lagged values of the
	corresponding price series form a high-dimensional dynamic feature space.
	PCA is subsequently applied to extract the dominant modes of temporal variability,
	thereby mitigating multicollinearity and overfitting effects inherent in high-dimensional
	lag representations.
	
	The coefficient of determination ($R^2$) is employed as the sole performance metric
	to ensure a consistent and interpretable comparison between training and test sets.
	For each asset, the number of retained principal components is varied systematically,
	and the corresponding $R^2$ values are recorded.
	
	\begin{figure}[H]
		\centering
		\includegraphics[width=0.95\linewidth]{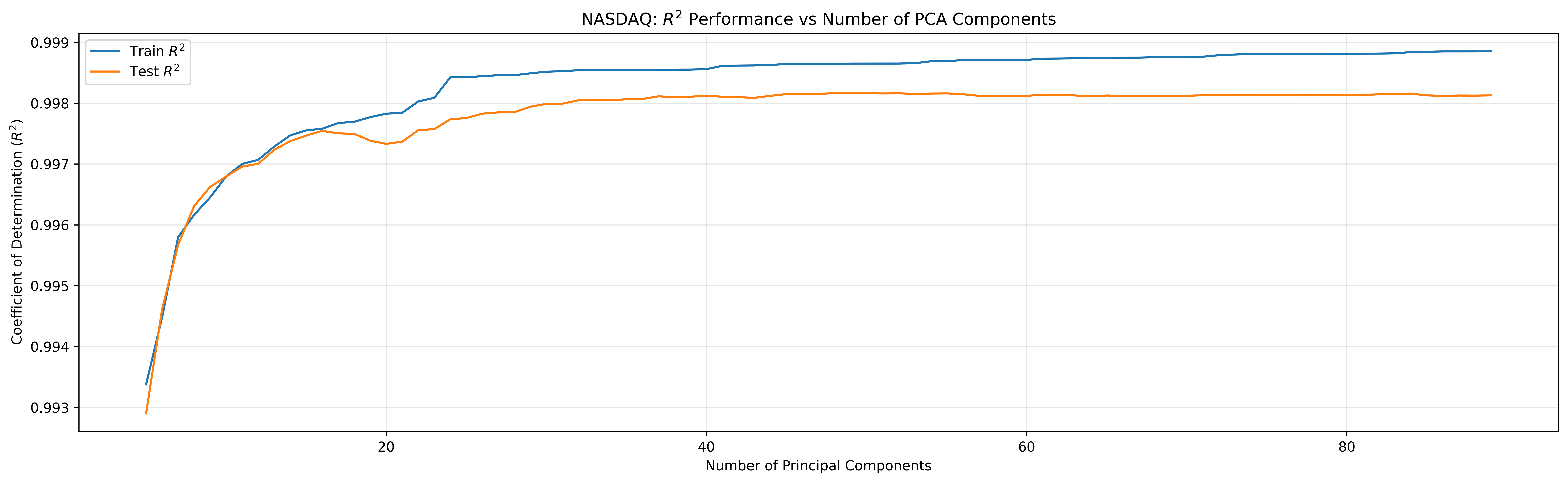}
		\caption{$R^2$ performance of the PCA--regression model for \textbf{Nasdaq}.
			The saturation of test performance indicates that most predictive information
			is captured by a limited number of dominant components.}
		\label{fig:pca_nasdaq}
	\end{figure}
	
	\begin{figure}[H]
		\centering
		\includegraphics[width=0.95\linewidth]{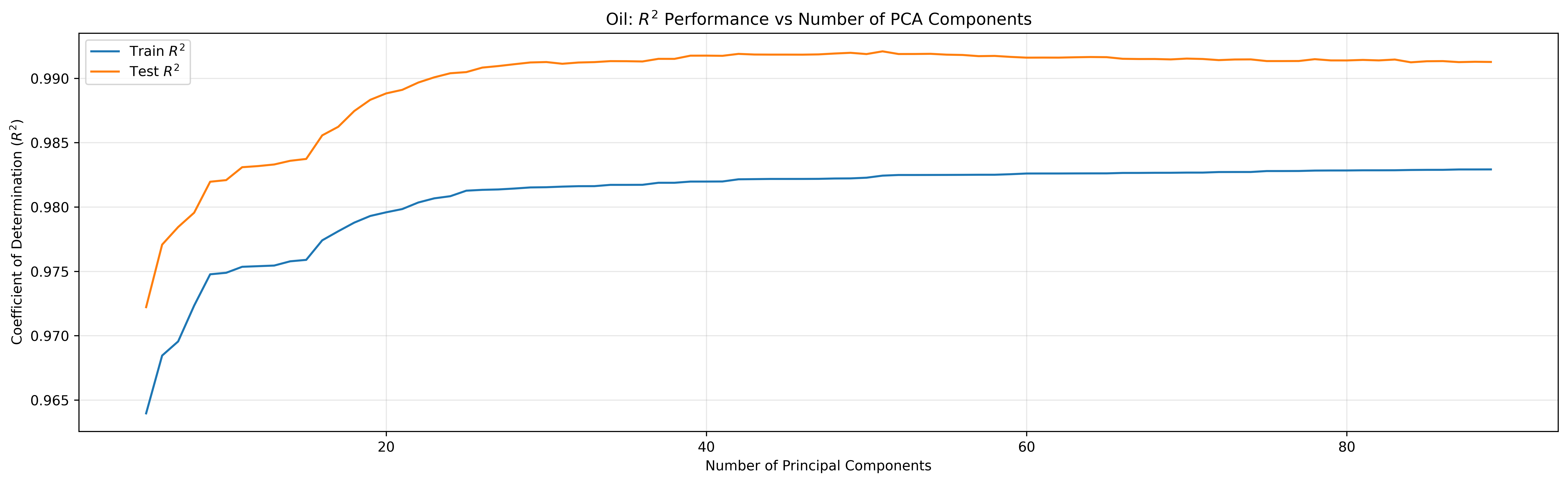}
		\caption{$R^2$ performance of the PCA--regression model for \textbf{Crude Oil}.
			Test accuracy improves steadily with increasing dimensionality before reaching
			a plateau, suggesting diminishing returns beyond the optimal subspace.}
		\label{fig:pca_oil}
	\end{figure}
	
	\begin{figure}[H]
		\centering
		\includegraphics[width=0.95\linewidth]{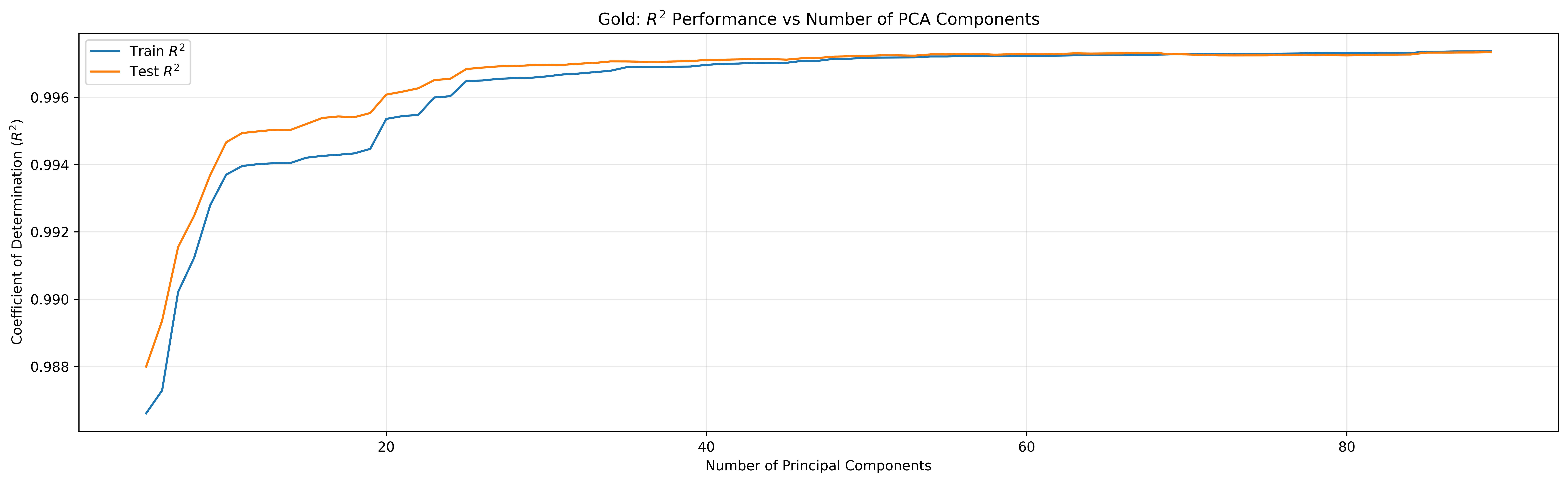}
		\caption{$R^2$ performance of the PCA--regression model for \textbf{Gold}.
			The close agreement between training and test curves indicates robust generalization
			and limited overfitting across a wide range of components.}
		\label{fig:pca_gold}
	\end{figure}
	
	\begin{figure}[H]
		\centering
		\includegraphics[width=0.95\linewidth]{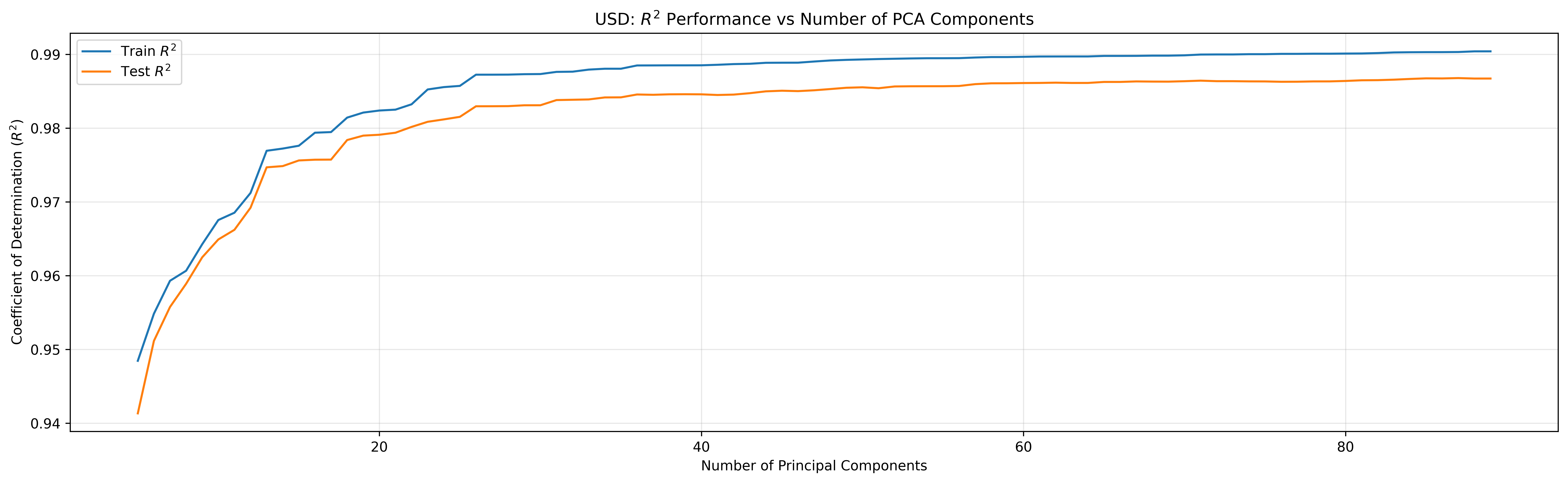}
		\caption{$R^2$ performance of the PCA--regression model for \textbf{USD}.
			Compared to other assets, the test $R^2$ exhibits a slower convergence, reflecting
			the more complex and distributed temporal structure of currency dynamics.}
		\label{fig:pca_usd}
	\end{figure}
	
	Across all assets, a clear saturation behavior is observed: after a certain number
	of principal components, the marginal improvement in test $R^2$ becomes negligible.
	This observation motivates the selection of an optimal dimensionality based on a
	penalized performance criterion, balancing accuracy and model complexity.
	
	\begin{table}[H]
		\centering
		\caption{Optimal number of PCA components and corresponding penalized test $R^2$
			for each financial asset.}
		\label{tab:pca_optimal}
		\begin{tabular}{lcccc}
			\hline
			\textbf{Asset} & \textbf{Nasdaq} & \textbf{Crude-Oil} & \textbf{Gold} & \textbf{USD} \\
			\hline
			Optimal PCA components & 49 & 51 & 89 & 87 \\
			Penalized Test $R^2$ & 0.9982 & 0.9921 & 0.9973 & 0.9868 \\
			\hline
		\end{tabular}
	\end{table}
	
	The results reported in Table~\ref{tab:pca_optimal} demonstrate that all four financial
	assets exhibit strong predictability when their dynamics are encoded through lagged
	representations and projected onto a suitably chosen low-dimensional subspace.
	Notably, equity and commodity markets (Nasdaq and Gold) achieve near-perfect explanatory
	power with fewer components compared to the USD market, whose dynamics appear to be
	more diffused across higher-order modes.
	
	Overall, these findings highlight the effectiveness of PCA as a dimensionality reduction
	tool for dynamically constructed financial datasets. By retaining only the most
	informative temporal components, the proposed framework achieves high predictive
	accuracy while preserving interpretability and numerical stability.

	\section{Conclusion}
	
	In this study, the dynamic interdependencies among four major financial assets, namely the Nasdaq index, crude oil, gold, and the U.S. dollar, were systematically investigated using a combination of qualitative economic reasoning and quantitative statistical modeling. The analysis was structured in two complementary parts, aiming to uncover both the structural relationships and the predictive behavior of these financial time series.
	
	In the first part of the Results and Discussion section, a qualitative assessment of cross-asset interactions was conducted based on established economic mechanisms and stylized facts. This analysis highlighted the prominent role of energy prices, safe-haven assets, and exchange rates in shaping equity market dynamics, while emphasizing the inherently nonlinear and interconnected nature of financial markets. The subsequent quantitative analysis, employing correlation measures and lagged multivariate linear regression models, provided empirical validation for these qualitative insights. In particular, the estimated regression coefficients and correlation structures consistently confirmed the presence of significant cross-asset effects, thereby demonstrating that the quantitative findings effectively satisfy and support the qualitative interpretations presented earlier.
	
	The second part of the analysis focused on predictive performance, where dynamically constructed lagged features were encoded and reduced using principal component analysis. Model performance was evaluated exclusively through the coefficient of determination ($R^2$), allowing for a clear and consistent comparison across assets. The results revealed that all four financial series exhibit strong predictability under an appropriate number of principal components, with distinct optimal dimensionalities reflecting differences in their underlying market structures. The observed saturation behavior of the test $R^2$ curves further indicates that excessive model complexity does not necessarily translate into improved generalization, highlighting the importance of balanced dimensionality selection.
	
	Overall, the findings demonstrate that combining economically motivated qualitative analysis with statistically grounded quantitative methods provides a coherent and robust framework for studying financial time series dynamics. The consistency between qualitative expectations and quantitative evidence strengthens the credibility of the proposed approach, while the strong predictive performance achieved across multiple assets underscores its practical relevance. These results suggest that the proposed framework can serve as a useful basis for future extensions toward nonlinear modeling, regime-dependent dynamics, or higher-frequency financial data.

\section*{Acknowledgement}

The authors acknowledge that an AI-based language model (ChatGPT, OpenAI) was used
solely to assist with English language editing. All scientific content, data analysis,
and interpretations were performed and verified by the authors.

	
	
	
	
	


\begin{thebibliography}{99}
		\bibitem{mantegna1999introduction}
		R. N. Mantegna and H. E. Stanley,
		\emph{"Introduction to Econophysics: Correlations and Complexity in Finance"},
		Cambridge University Press (1999).
		
		\bibitem{laloux1999noise}
		L. Laloux, P. Cizeau, J.-P. Bouchaud, and M. Potters,
		\emph{"Noise dressing of financial correlation matrices"},
		Physical Review Letters 83, 1467 (1999).
		
		\bibitem{bouchaud2003theory}
		J.-P. Bouchaud and M. Potters,
		\emph{"Theory of Financial Risk and Derivative Pricing"},
		Cambridge University Press (2003).
		
		\bibitem{brunton2016discovering}
		S. L. Brunton, J. L. Proctor, and J. N. Kutz,
		\emph{"Discovering governing equations from data by sparse identification of nonlinear dynamical systems"},
		Proceedings of the National Academy of Sciences 113 (15), 3932–3937 (2016).
		
		\bibitem{takens1981detecting}
		F. Takens,
		\emph{"Detecting strange attractors in turbulence"},
		Lecture Notes in Mathematics 898, Springer (1981).
		
		\bibitem{kantz2004nonlinear}
		H. Kantz and T. Schreiber,
		\emph{"Nonlinear Time Series Analysis"},
		Cambridge University Press (2004).
		
		\bibitem{alados2014positive}
		C. Alados, P. Errea et al.,
		\emph{"Positive and negative feedbacks and free-scale pattern distribution in rural-population dynamics"},
		PLOS One 9 (12), e114561 (2014).
		
		\bibitem{khalilian2025mapping}
		P. Khalilian, A. N. Golestani, M. Eslamifar, and M. T. Firouzjaee,
		\emph{"Mapping Crisis-Driven Market Dynamics: A Transfer Entropy and Kramers-Moyal Approach to Financial Networks"},
		arXiv:2507.09554 [q-fin.ST] (2025). :contentReference[oaicite:0]{index=0}
		
		
		\bibitem{salisu2017revisiting} A. A Salisu, K. O Isah, \emph{"Revisiting the oil price and stock market nexus: A nonlinear Panel ARDL approach"}. Econ. Model., Vol. 66, Pages 258-271 (2017).
		
		\bibitem{su2021does} C.-W. Su, T. Sun et al., \emph{"Does institutional quality and remittances inflow crowd-in private investment to avoid Dutch Disease? A case for emerging seven (E7) economies"}. Resour. Pol., Vol. 72 102111 (2021).
		
		\bibitem{jianqiang2020exacerbating} J. GU, M. Umar et al., \emph{"Exacerbating effect of energy prices on resource curse: Can research and development be a mitigating factor?"}. Resour. Pol., Vol. 67 101689 (2020).
		
		\bibitem{naseer2020double} A. Naseera, C.-W, Su et al., \emph{"Double jeopardy of resources and investment curse in South Asia: Is technology the only way out?"}. Resour. Pol., Vol. 68 101702 (2020).
		
		\bibitem{guan2021volatility} L. Guana, W. -W, Zhanga et al., \emph{"The volatility of natural resource prices and its impact on the economic growth for natural resource-dependent economies: A comparison of oil and gold dependent economies"}. Resour. Pol., Vol. 72 102125 (2021).
		
		\bibitem{lucey2014gold} B. M Lucey, C. Larkin et al., \emph{"Gold markets around the world--who spills over what, to whom, when?"}. Appl. Econ. Lett., Vol. 21, 887-892  (2014).
		
		\bibitem{arfaoui2017oil} M. Arfaoui, A. B. Rejeb, \emph{"Oil, Gold, US dollar and Stock market interdependencies: A global analytical insight"}. European Journal of Management and Business Economics, Vol. 26 No. 3, pp. 278-293 (2017).
		\bibitem{golub1983oil} S. Golub \emph{"Oil prices and exchange rates"}. The Economic Journal 93, 576–593 (1983).
		\bibitem{krugman1983oil} P. Krugman, \emph{"Oil shocks and exchange rate dynamics"}. in F. J. A., ed., ―Exchange rates and international macroeconomics, University of Chicago Press. (1983).
		\bibitem{karasu2022crude} S. Karasu, A. Altan, \emph{"Crude oil time series prediction model based on LSTM network with chaotic Henry gas solubility optimization"}. Energy 242(7) (2021).
		\bibitem{ling2011chaotic} L. -Y. He \emph{"Chaotic Structures in Brent and WTI Crude Oil Markets: Empirical Evidence"}. Int. J. Econ. Financ., 3,
		242–249 (2011).
		\bibitem{sujit2011study} K. S. Sujit, B. R. Kumar \emph{"Study on dynamic relationship among gold price, oil price, exchange rate and stock market returns"}. IJBESAR 9(2):145-165 (2011).
		\bibitem{an2019oil} J. An, A. Mikhaylov et al. \emph{"Oil price predictors: Machine learning approach"}. , IJEEP, 9(5), 1-6 (2019).
		\bibitem{nyangarika2019oil} A. Nyangarika1, A. Mikhaylov et al. \emph{"Oil Price Factors: Forecasting on the Base of Modified Auto-regressive Integrated Moving Average Model"}. , IJEEP, 9(1), 149-159 (2019).
		\bibitem{miao2017influential} H. Miao, S. Ramchander \emph{"Influential factors in crude oil price forecasting"}. Energy Econ. 68 77–88 (2017).
		\bibitem{novotny2012link} F. Novotný \emph{"The link between the Brent crude oil price and the US dollar exchange rate"}. , Prague Econ. Pap., 21.2 220-232 (2012).
		\bibitem{dees2008assessing} S. Dées, A. Gasteuil et al. \emph{"Assessing the factors behind oil price changes"}. , No. 855. ECB working paper series (2008).
		\bibitem{shen2014us} Z. Shen \emph{"How the US dollar index affects gold prices"}., A research project submitted in partial fulfillment of the requirements for the degree of Master of Finance. Saint Mary’s University: Zhengyuan Shen (2014).
		\bibitem{juarez2011applying} F. Juárez \emph{"Applying the theory of chaos and a complex model of health to establish relations among financial indicators"}., Procedia Comput. Sci. 3, 982-986 (2014).
		\bibitem{oloko2021fractional} T. F.Oloko, A. E.Ogbonna et al. \emph{"Fractional cointegration between gold price and inflation rate: implication for inflation rate persistence"}., Resour. Policy, 74, 102369 (2021).
		\bibitem{tkacz2007gold} G. Tkacz \emph{"Gold prices and inflation"}., No. 2007, 35. Bank of Canada Working Paper (2007).
		\bibitem{firouzjaee2022machine} J. T. Firouzjaee, P. Khaliliyan \emph{"Machine learning model to project the impact of Ukraine crisis"}., arXiv preprint arXiv:2203.01738 (2022).
		
		\bibitem{pagano2008quality} M. S. Pagano, L. Peng et al. \emph{"The quality of price formation at market openings and closings: Evidence from the NASDAQ stock market"}. , No. 2008/45. CFS Working Paper (2008).
		\bibitem{pagano2013call} M. S. Pagano, L. Peng et al. \emph{"A call auction’s impact on price formation and order routing: evidence from the NASDAQ stock market"}. , J. Financ. Mark., 16(2), 331-361 (2013).
		\bibitem{ellul2005opening} A. Ellul, H. S. Shin \emph{"Opening and Closing the Market: Evidence from the London Stock Exchange"}., JFQA, 40(4), 779-801 (2005).
		\bibitem{pagano1989trading} M. Pagano \emph{"Trading volume and asset liquidity"}., Q. J. Econ. 104, 255– 274 (1989).
		\bibitem{chakraborty2012order} A. Chakraborty, M.S. Pagano et al. \emph{"Order revelation at market openings"}., J. Financ. Mark., 15, 127–150 (2012).
		\bibitem{tibshirani1996regression} R. Tibshirani \emph{"Regression shrinkage and selection via the LASSO"}. J. R. Stat. Soc. Ser. 58, 267–288 (1996).
		\bibitem{osborne2000lasso} M.R. Osborne, B. Presnell et al. \emph{"On the LASSO and its dual"}. J. Comput. Graph. Stat. 9, 319–337 (2000).
		\bibitem{ma2018forecasting} F. Ma, J. Liu et al. \emph{"Forecasting the aggregate oil price volatility in a data-rich environment"}. Econ. Model.,  72, Pages 320-332 (2018).
		\bibitem{zhang2019forecasting} Y. Zhang, F. Ma et al. \emph{"Forecasting crude oil prices with a large set of predictors: Can LASSO select powerful predictors?"}., J. Empir. Finance, 54, Pages 97-117 (2019).
		\bibitem{zare2018extension} A. Zare, A. Ozdemir et al. \emph{"Extension of PCA to higher order data structures: An introduction to tensors, tensor decompositions, and tensor PCA"}., Proc. IEEE, 106(8), 1341-1358 (2018).
		\bibitem{cunningham2015linear} J. P. Cunningham, Z. Ghahramani \emph{"Linear dimensionality reduction: Survey, insights, and generalizations"}., JMLR, 16(1), 2859-2900 (2015).
		\bibitem{jiang2011linear} X. Jiang \emph{"Linear subspace learning-based dimensionality reduction"}., IEEE Signal Process. Mag., 28(2), 16-26 (2011).
		\bibitem{candes2011robust} E. J. Candes, X. Li et al. \emph{"Robust principal component analysis"}., J. ACM, 58(3), 1-37 (2011).
		\bibitem{wright2009robust} J. Wright, A. Ganesh et al. \emph{"Robust principal component analysis: Exact recovery of corrupted low-rank matrices via convex optimization"}., Adv. Neural Inf. Process Syst. , 22 (2009).
		\bibitem{francis1999introduction} P. J. Francis, B. J. Wills \emph{"Introduction to Principal Components Analysis"}.,  arXiv preprint astro-ph/9905079 (1999).
		\bibitem{kazem2013support} A. Kazem \emph{"Support vector regression with chaos-based firefly algorithm for stock market price forecasting"}. Appl. Soft Comput, 13(2), 947-958 (2013).
		\bibitem{qi2020event} L. Qi, M. Khushi et al. \emph{"Event-driven lstm for forex price prediction"}. In 2020 IEEE Asia-Pacific Conference on Computer Science and Data Engineering (CSDE), pp. 1-6. IEEE (2020).
		\bibitem{park2013stock} K. Park, H. Shin \emph{"Stock price prediction based on a complex interrelation network of economic factors"}. Eng. Appl. Artif. Intell.,  26(5-6), 1550-1561 (2013).

		\bibitem{firouzjaee2022lstm} Firouzjaee, J.T. and Khalilian, P., 2024. The interpretability of LSTM models for predicting oil company stocks: impact of correlated features. International Journal of Energy Research, 2024(1), p.5526692.
		






	
		
		
		
		
		
		
		
	\end{thebibliography}
\end{document}